\documentclass[12pt,draftclsnofoot,onecolumn]{IEEEtran}

\usepackage{enumitem}
\usepackage{amsmath,graphics,amssymb,epsfig,subfigure,color,cite}
\usepackage{array}
\usepackage{multirow}
\usepackage{enumerate}
\usepackage{algorithm2e}
\usepackage{algpseudocode}
\usepackage{algorithmicx}
\usepackage{algcompatible}
\usepackage{bm}
\usepackage{float}
\usepackage{makecell}
\usepackage[right=0.625in, left=0.625in, top=0.75in, bottom=1.0in]{geometry}

\makeatletter
\newcommand{\vast}{\bBigg@{4.5}}
\newcommand{\Vast}{\bBigg@{7.5}}
\makeatother

\begin{document}
	\title{Vector Quantization for Deep-Learning-Based CSI Feedback in Massive MIMO Systems}

	\author{Junyong Shin, Yujin Kang, and Yo-Seb Jeon 
		\thanks{Junyong Shin, Yujin Kang, and Yo-Seb Jeon are with the Department of Electrical Engineering, POSTECH, Pohang, Gyeongbuk, Republic of Korea  (e-mail: sjyong@postech.ac.kr, yujinkang@postech.ac.kr, yoseb.jeon@postech.ac.kr).} 
	}
\vspace{-2mm}

\maketitle
\vspace{-12mm}

\begin{abstract} 
	This paper presents a finite-rate deep-learning (DL)-based channel state information (CSI) feedback method for massive multiple-input multiple-output (MIMO) systems. The presented method provides a finite-bit representation of the latent vector based on a vector-quantized variational autoencoder (VQ-VAE) framework while reducing its computational complexity based on shape-gain vector quantization. In this method, the magnitude of the latent vector is quantized using a non-uniform scalar codebook with a proper transformation function, while the direction of the latent vector is quantized using a trainable Grassmannian codebook. A multi-rate codebook design strategy is also developed by introducing a codeword selection rule for a nested codebook along with the design of a loss function. Simulation results demonstrate that the proposed method reduces the computational complexity associated with VQ-VAE while improving CSI reconstruction performance under a given feedback overhead. 
\end{abstract}


\section{Introduction}\label{Sec:Intro}

Massive multiple-input multiple-output (MIMO) stands out as a crucial technique for enhancing spectral efficiency in wireless communication systems. In frequency division duplexing (FDD) systems, achieving this improvement necessitates accurate knowledge of the channel state information (CSI) at the base station (BS). However, as the dimension of CSI grows substantially in massive MIMO systems, there emerges a significant burden on user equipment (UE) to feed the CSI back to a BS. To address this challenge, CSI feedback methods based on recent advancements in deep learning (DL) have been introduced \cite{CsiNet,CRNet,CsiNet+}. 
The common idea of the DL-based CSI feedback methods, which is based on an autoencoder (AE) framework, is to compress the CSI using an encoder network and then feed the latent vector (i.e., the output of the encoder) back to the BS with reduced overheads. 
With appropriate pre-processing of the CSI and meticulous design of the neural networks, the above  methods have proven effective in mitigating the CSI feedback overhead, compared to conventional non-DL-based approaches such as compressive sensing and static codebook-based feedback\cite{conventional1, conventional2}.

In practice, for compatibility with modern digital communication systems, the latent vector, corresponding to the UE's feedback, needs to be transformed into a finite-length bit sequence. 
Motivated by this fact, scalar quantization methods for the DL-based CSI feedback have been developed to support the finite-bit representation of the latent vector. 
In \cite{CsiNet+}, non-uniform scalar quantizer and dequantization module were developed. In this method, the entries of the latent vector were normalized according to their distributions, so that these entries can be quantized in a bounded range. 
In \cite{CQNet, Magnitude Quantization}, bounded activation functions were considered in order to use a non-uniform quantizer in a bounded range.
All these methods assume independent quantization of the latent entries. This assumption, however, making them impossible to leverage the correlations among different latent entries, despite its potential to reduce quantization error as mentioned in \cite{VQpros.}.
Another limitation is that when employing scalar quantization, at least one bit is required for quantizing each entry. This requirement significantly restricts the dimension of the latent vector under a given feedback overhead, resulting in performance degradation.
The limitations of the scalar-quantization approach have been addressed in \cite{OVQ,EfficientFi} by incorporating the idea of vector quantization into the DL-based CSI feedback. 
The common strategy of these methods is to leverage the vector-quantized variational autoencoder (VQ-VAE) framework developed in \cite{VQ-VAE} which allows a finite-bit representation of the latent vector based on a trainable vector codebook. 
However, these methods require comparing the distances from the latent vector to all codeword vectors, resulting in significant computational complexity that scales with the size of the codebook. Moreover, a vector codebook design for DL-based CSI feedback has not been explored in the literature, despite its importance and potential for minimizing quantization error.   



In this letter, we propose a finite-rate DL-based CSI feedback method for massive MIMO systems, utilizing the VQ-VAE framework.
Our key contribution is to design the codebook for VQ-VAE based on shape-gain vector quantization, which alleviates the computational complexity associated with VQ-VAE by separately quantizing the shape (direction) and gain (magnitude) of the latent vector.
In particular, for the gain quantization, we adopt a non-uniform scalar codebook based on a clipped $\mu$-law transformation which captures the behavior of the latent vector's magnitude. For the shape quantization, we employ a trainable Grassmannian codebook under the unit-norm constraint.
Beyond the design of a single-rate vector codebook, we also develop a multi-rate codebook design strategy for the proposed method to support multi-rate vector quantization using a single codebook. In this strategy, we modify a nested codebook design in \cite{Nested Quantization}, by developing a new codeword selection rule along with a revised loss function design.
In simulations, we demonstrate that the proposed method significantly reduces the computational complexity of the quantization process while improving reconstruction performance under a given feedback overhead.  
Our simulation results also demonstrate that the use of our multi-rate codebook design further improves the performance of the proposed method.

\section{System Model and Preliminary}\label{Sec:Model}

\subsection{FDD Massive MIMO System}
Consider a single-cell massive MIMO system in which a BS equipped with $N_t$ transmit antennas communicates with a UE equipped with a single antenna. The system employs orthogonal frequency division multiplexing (OFDM) with $N_c$ subcarriers. CSI in the spatial-frequency domain is represented as a channel matrix $\mathbf{H}_{\text{sf}}\in\mathbb{C}^{N_c \times N_t}$. To exploit the sparsity of the CSI matrix in the angular-delay domain as in \cite{CsiNet}, it is transformed into a channel matrix $\mathbf{H}_{\text{ad}}\in\mathbb{C}^{N_c \times N_t}$ by applying a 2D discrete Fourier transform (DFT). The transformation is formulated as $\mathbf{H}_{\text{ad}} = \mathbf{F}_\text d\mathbf{H}_{\text{sf}}\mathbf{F}_\text a,$
where $\mathbf{F}_\text d$ and $\mathbf{F}_\text a$ are $N_c \times N_c$ and $N_t \times N_t$ DFT matrices, respectively. In the angular-delay domain, only the first $\tilde {N}_c (\leq N_c)$ delay components carry significant values because multipath delays are confined to a limited time interval. Motivated by this, a submatrix of $\mathbf{H}_{\text{ad}}$ which consists of the initial $\tilde N_c$ rows is defined as an effective channel matrix $\tilde {\mathbf{H}}_{\text{ad}}\in \mathbb{C}^{\tilde N_c \times N_t}$. 

In a typical DL-based CSI feedback method, the UE employs an encoder network to compress the CSI information into the form of a latent vector with  dimension $M$. 
Meanwhile, the BS employs a decoder network to reconstruct the CSI information from the UE's feedback. 
The UE initiates a CSI feedback process by utilizing the effective channel matrix $\tilde {\mathbf{H}}_{\text{ad}}$  as an input of the encoder network.
This yields the latent vector ${\bf z}$ expressed as 
${\bf z} = f_{\text{enc}}(\tilde {\mathbf{H}}_{\text{ad}})$, where $f_{\text{enc}}$ represents the encoder network. Typically, the latent vector ${\bf z}$ is considered as a UE's feedback that needs to be transmitted to the BS.
Upon the reception of the UE's feedback, the BS reconstructs  the effective channel matrix by utilizing the latent vector as an input of the decoder network $f_{\text{dec}}$. 

In practice, the latent vector ${\bf z}$ is transformed into a finite-length bit sequence before being transmitted to the BS. This transformation is achieved through quantizing the latent vector $\mathbf z$ under the constraint on feedback overhead. Subsequently, the BS reconstructs the effective channel matrix by utilizing the quantized latent vector, denoted as ${\bf z}_q$, as an input for the decoder network (i.e., $\mathbf{\hat H} = f_{\text{dec}}(\mathbf{z}_q)$).



\subsection{Vector-Quantized Variational Autoencoder (VQ-VAE)} 
VQ-VAE is well-known for its ability to enable a discrete representation of the latent vector by incorporating vector quantization into the VAE framework \cite{VQ-VAE}. This framework utilizes a trainable quantization codebook placed in the latent space and jointly trains the encoder, codebook, and decoder using a loss function that captures both the quantization and reconstruction errors.

A conventional way of employing the VQ-VAE is to divide the latent vector ${\bf z}$ into $N$ sub-vectors each with dimension $D$ ($M=N\times D$) and then use a $D$-dimensional codebook for quantizing each sub-vector separately \cite{VQ-VAE,OVQ,EfficientFi}.
Let $\mathcal{B}$ be a vector codebook using $B$ bits which consists of $2^B$ $D$-dimensional codewords, namely $\{\mathbf{b}_k\}^{2^B}_{k=1}$. 
Also, let ${\bf z}_i$ be the $i$-th sub-vector of ${\bf z}$, defined as ${\bf z}_i = [z_{(i-1)D+1},\cdots,z_{i D}]$, where $z_j$ is the $j$-th entry of ${\bf z}$.
Then each sub-vector ${\bf z}_i$ is quantized to $\mathbf{z}_{q,i}$ using the codebook $\mathcal{B}$ according to the minimum Euclidean distance criterion, i.e., $\mathbf{z}_{q,i}=\text{argmin}_{\mathbf{b}_k \in \mathcal{B}}\Vert\mathbf{z}_i - \mathbf{b}_k \Vert$. 
For jointly training the encoder, codebook, and decoder, in \cite{VQ-VAE}, the loss function is designed as 
\begin{align}\label{eq: VQ-VAE Loss}
	\mathcal{L}_{\text{vq}} = \big\Vert\mathbf{\hat H} - \tilde {\mathbf{H}}_{\text{ad}} \big\Vert^2_{\text F}+\left\Vert \text{sg}(\mathbf z)-\mathbf{z}_q\right\Vert^2 + \beta \left\Vert\mathbf z-\text{sg}(\mathbf{z}_q)\right\Vert^2,
\end{align}
where $\text{sg}(\cdot)$ denotes the stop-gradient operator that ignores gradient descent computation as a constant. The third term in \eqref{eq: VQ-VAE Loss} is called as {\em commitment loss}, regularized by a hyperparameter $\beta$ \cite{VQ-VAE}. After calculating the quantization errors, which are the second and third term above, a gradient correction of the decoder input is performed as $\mathbf{z}_q\leftarrow\mathbf{z}+\text{sg}(\mathbf{z}_q-\mathbf{z})$. 

Unlike scalar quantization which ignores the correlations among the latent entries (e.g., \cite{CsiNet+, Magnitude Quantization,CQNet}), the VQ-VAE approach is able to capture joint behavior of multiple latent entries and therefore has a potential to achieve a further reduction in the quantization error under the same feedback overhead. However, the above VQ-VAE approach often imposes significant computational complexity because it compares the distances from the latent vector to all codeword vectors. This limitation poses a potential obstacle to the practical adoption of VQ-VAE as a solution for finite-rate DL-based CSI feedback. 

\section{Proposed DL-based CSI Feedback Method}
In this section, we propose a novel DL-based CSI feedback method which reduces the computational complexity of the original VQ-VAE approach by leveraging shape-gain vector quantization. 

\subsection{Basic Idea: Shape-Gain Quantization}
\begin{figure}[t]
	\centering 
	{\epsfig{file=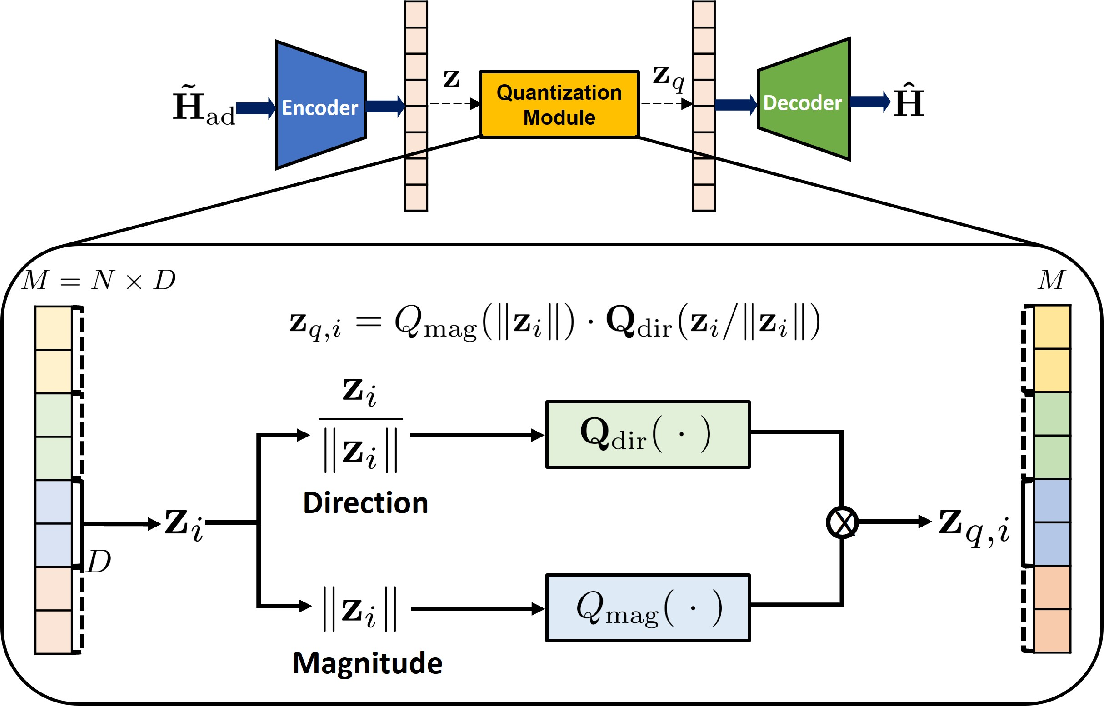, width=10cm}} 
	\caption{An illustration of the proposed DL-based CSI feedback method using shape-gain vector quantization.} 
	\label{fig:TVQ_structure}\vspace{-3mm}
\end{figure}

The basic idea of the proposed method is to quantize the magnitude and direction of each latent sub-vector ${\bf z}_i$ using a gain and shape quantizer, respectively. 
To be more specific, we quantize the magnitude $\|{\bf z}_i\|$ of ${\bf z}_i$ using a gain quantizer $Q_{\rm mag}(\cdot)$, while quantizing the direction ${\bf z}_i/\|{\bf z}_i\|$ of ${\bf z}_i$  using a shape quantizer ${\bf Q}_{\rm dir}$. Then, the quantized latent sub-vector is given by 
\begin{align}\label{eq: Two-stage Quantization}
	\mathbf{z}_{q,i} = Q_{\text{mag}}(\Vert \mathbf{z}_i \Vert) \cdot \mathbf{Q}_{\text{dir}}(\mathbf{z}_i/\Vert \mathbf{z}_i \Vert).
\end{align}
Our shape-gain quantization strategy is illustrated in Fig. \ref{fig:TVQ_structure}.


A significant advantage of our quantization strategy is the reduced computational complexity in the quantization process. In our approach, the feedback bits per sub-vector are divided between the shape and gain quantizers. Let $B_{\rm mag}$ and $B_{\rm dir}$ denote the feedback bits for the gain and shape quantizers, respectively, implying that $B_{\rm mag} + B_{\rm dir} = B$, when the feedback overhead per sub-vector is $B$ bits. It leads for the entire feedback overheads to be calculated as $\frac{M}{D}\times B$. In this case, the computational complexity of our shape-gain quantization in \eqref{eq: Two-stage Quantization} is of order $\mathcal{O}(2^{\max\{B_{\rm mag},B_{\rm dir}\}})$, significantly smaller than the complexity order $\mathcal{O}(2^B)$ of the original VQ-VAE approach. Therefore, for the same feedback overhead, our quantization approach significantly reduces the computational complexity compared to the VQ-VAE approach.



\subsection{Gain (Magnitude) Quantization}
We present the design of the gain quantizer $Q_{\text{mag}}(\cdot)$ used to quantize the magnitude $\|{\bf z}_i\|$ of each latent sub-vector ${\bf z}_i$. It should first be noted that each latent sub-vector ${\bf z}_i$ has a bounded magnitude if the activation function of the output layer of the encoder is bounded. Without loss of generality, we shall assume that this output function is bounded by $1$. Consequently, the magnitude $\|{\bf z}_i\|$ is constrained within $\sqrt{D}$. Another crucial observation is that the magnitude of the sub-vector is often concentrated in a specific regime rather than evenly distributed across the range of $[0, \sqrt{D}]$. This observation suggests that a simple uniform quantizer may not perform optimally for the quantization of $\|{\bf z}_i\|$.
Motivated by this observation, we construct a non-uniform quantizer by mapping the magnitude $\|{\bf z}_i\|$ to the range of $[0,A]$ with an appropriate transformation $h(x): [0, \sqrt{D}] \rightarrow [0,A]$ and perform uniform quantization as \eqref{eq: clipped uniform}. 
Our gain quantizer is then expressed as $Q_{\text{mag}}(\Vert \mathbf{z}_i \Vert) = h^{-1}(f_{u}(h(\Vert \mathbf{z}_i \Vert)))$, where $f_{u}(x)$ is a $B_{\rm mag}$-bit uniform quantization function defined as
\begin{align}\label{eq: clipped uniform}
	f_{u}(x) = A\left\{\frac{\text{round}(2^{B_{\rm mag}}{x}/{A}-0.5)+0.5}{2^{B_{\rm mag}}}\right\}.
\end{align}

Next, for determining an appropriate transformation $h(x)$, we consider the $\mu$-law transformation expressed as $h_{\mu\text{-law}}(x) = \text{ln}\big(1 + \frac{\mu x}{\sqrt{D}}\big)/\text{ln}(1 + \mu)$, which has been widely studied in the literature \cite{CsiNet+, Magnitude Quantization, CQNet}. This method provides concentrated quantization levels near-zero values due to its concavity for positive inputs. To restrict the output of this method to focus only on a confined region $[0,A]$, we modify the original $\mu$-law transformation by introducing a clipping value, which limits the highest value for the output of the transformation. Our clipped $\mu$-law transformation is expressed as follows:
	\begin{align}\label{eq: clipped mu-law}
		\hat h_{\mu\text{-law}}(x)=\begin{cases}h_{\mu\text{-law}}(x),& 0\leq {x} \leq \frac{(1+\mu)^A-1}{\mu}\sqrt{D},\\
			A,& \frac{(1+\mu)^A-1}{\mu}\sqrt{D}< {x} \leq \sqrt{D}. \end{cases}
	\end{align}

	When training the encoder and decoder, our gain quantizer does not allow a direct back propagation for gradient computing, because the round functions in \eqref{eq: clipped uniform} have a zero-gradient problem. To resolve this problem, 
	we leverage a soft-gradient passing technique considered in \cite{Magnitude Quantization, Magnitude Quantization2}. When executing the back propagation process, we replace round functions in \eqref{eq: clipped uniform}  with the step-wise Tanh functions as follows:
	\begin{align}\label{eq: soft gradient}
		\tilde{f}_{u}(x) = 
		\frac{A}{2}\left\{\frac{\sum^{2^{B_{\rm mag}-1}}_{i=1}\text{tanh}(\tau(2^{B_{\rm mag}}x/A-i))+1}{2^{B_{\rm mag}}}\right\}.
	\end{align}

	\subsection{Shape (Direction) Quantization}
	We introduce the design of the shape quantizer ${\bf Q}_{\rm dir}(\cdot)$ used to quantize the direction $\tilde{\mathbf{z}}_i  = {\bf z}_i/\|{\bf z}_i\|$ of each latent sub-vector ${\bf z}_i$. 
	Note that a distance between two unit-norm vectors ${\bf c}_1$ and ${\bf c}_2$ can be measured by the sine of the angle between these vectors, defined as $d({\bf c}_1,{\bf c}_2) = \sqrt{1-|{\bf c}_1,{\bf c}_2|^2}$. 
	Utilizing this fact, we adopt the quantization function that determines the codeword with the minimum distance according to the above distance measure: 
	\begin{align}
		\mathbf{Q}_{\text{dir}}(\tilde{\mathbf{z}}_i) = \underset{\mathbf{b}_k \in \mathcal{B}_{\rm dir}}{\arg\!\min}  ~d(\tilde{\mathbf{z}}_i,\mathbf{b}_k),
	\end{align}
	where $\mathcal{B}_{\rm dir}$ is a shape codebook which consists of $2^{B_{\rm dir}}$ unit-norm vectors with dimension $D$. 
	
	We now focus on designing a proper shape codebook. At the beginning of a training process, we do not have any prior information about the direction of the latent sub-vectors. Hence we initialize the codebook by assuming that the directions of the latent sub-vectors are uniformly distributed. Under this assumption, minimizing the quantization error of our shape quantizer is achieved by determining the $2^{B_{\rm dir}}$ codewords that maximize the minimum distance between any two codewords. 
	This problem is well known as a  {\em Grassmannian line packing} problem and can be solved efficiently by some existing algorithms \cite{Grass_LBG}. Utilizing this fact, we initialize the codebook $\mathcal{B}_{\rm dir}$ with the solution of the above problem, known as the {\em Grassmannian} codebook. As the training progresses, more information about the directions of the latent sub-vectors becomes available. Therefore, during the training process, we update the codewords in $\mathcal{B}_{\rm dir}$ based on the gradient with respect to the loss function. It's noteworthy that the vectors in the codebook may not have unit norm after a gradient descent update; therefore, we normalize the codebook vectors  after their updates.

	\subsection{Multi-Rate Codebook Design}\label{Sec:Multi}
	We also present a multi-rate codebook design strategy for the proposed CSI feedback method.
	A multi-rate VQ-VAE approach was studied in \cite{Nested Quantization} for image classification tasks. In this approach, a nested codebook is constructed by successively increasing the codebook size through the addition of random codeword vectors. We modify this strategy for designing a nested shape codebook to support $L$ different shape quantization rates. Unlike the existing approach in \cite{Nested Quantization}, our strategy involves constructing a nested shape codebook by successively reducing the codebook size from a large size to a small size. The resulting $L$ shape codebooks can be expressed as $\mathcal{B}_{\rm dir}^{(1)}\supset\mathcal{B}_{\rm dir}^{(2)}\supset\cdots \supset\mathcal{B}_{\rm dir}^{(L)}$ with $|\mathcal{B}_{\rm dir}^{(l)}| = 2^{B_{\rm dir}^{(l)}}$ for $l\in \{1,...,L\}$. The first codebook $\mathcal{B}_{\rm dir}^{(1)}$ is initialized as our shape-gain vector codebook and then trained using the loss function in \eqref{eq: VQ-VAE Loss}, as described earlier. Unlike the first codebook, the $l$-th codebook $\mathcal{B}^{(l)}$ for $l>1$ is initialized by selecting the most frequently quantized $2^{B^{(l)}_{\text{dir}}}$ vectors from the $(l-1)$-th codebook $\mathcal{B}^{(l-1)}$. This strategy represents an important deviation from the approach in \cite{Nested Quantization}, as our strategy involves selecting vectors based on their frequency of quantization rather than simply adding random codeword vectors. Then, an initialized $l$-th codebook is trained using the following loss function:
	\begin{align}\label{eq: nested quantization}
		\mathcal{L}^{(l)}_{\text{nvq}} = \frac{1}{\sum_{k=1}^{l}\gamma^k}\sum_{j=1}^{l}\gamma^j\Big(\big\Vert\mathbf{\hat H}^{(j)} - \tilde {\mathbf{H}}_{\text{ad}} \big\Vert^2_{\text F}+\Vert \text{sg}(\mathbf z)-\mathbf{z}^{(j)}_q\Vert^2 + \beta \Vert\mathbf z-\text{sg}(\mathbf{z}_q^{(j)})\Vert^2\Big),
	\end{align}
	where $\mathbf{z}^{(j)}_q$ is the latent vector quantized with $\mathcal{B}^{(j)}$, $\mathbf{\hat H}^{(j)} = f_{\text{dec}}(\mathbf{z}_q^{(j)})$, and  $\gamma\leq 1$ is a hyperparameter which regulates the portion of each quantization order's loss in the nested codebook. 
	Unlike the loss function in \eqref{eq: nested quantization} which considers a simple summation of the losses of $l$ codebooks, we consider a {\em weighted} combination of the losses of $l$ codebooks by assigning a smaller weight to a codebook with a larger size. By doing so, the $l$-th codebook $\mathcal{B}^{(l)}$ is not only optimized for the $l$-th rate, but also regulated to preserve its effectiveness in higher rates. 
	Our multi-rate codebook design strategy is summarized in Algorithm \ref{alg: multi-rate}. Similarly to other DL-based CSI feedback methods, when the channel distribution undergoes significant changes from those assumed during offline training, our shape-gain codebooks can be updated jointly with the encoder/decoder weights as part of the re-training or fine-tuning process.

	
	{\small \RestyleAlgo{ruled}
		\SetKwComment{Comment}{/* }{ */}
		\begin{algorithm}[t]\label{alg: multi-rate}
			\caption{Training of the proposed CSI feedback method with multi-rate codebook design}
			\textbf{1. Initialization:}\\
			Randomly initialize the parameters in $f_{\text{enc}}$ and $f_{\text{dec}}$;\\
			Initialize $\mathcal{B}_{\text{dir}}^{(1)}$ with the \textit{Grassmannian} codebook;\\
			\textbf{2. Training:}\\
			\For{$l=1,...,L$}{
				\While{not converged}{
					$\mathbf{z} \gets f_{\text{enc}}(\tilde{\mathbf{H}}_{\text{ad}})$;\\
					\For{$i=1,...,N$}{
						$Q_{\text{mag}} \gets h^{-1}_{\mu\text{-law}}(f_{u}(\hat{h}_{\mu\text{-law}}(\Vert \mathbf{z}_i \Vert)))$;\\
						$\nabla Q_{\text{mag}} \gets \nabla h^{-1}_{\mu\text{-law}}(\tilde{f}_{u}(\hat{h}_{\mu\text{-law}}(\Vert \mathbf{z}_i \Vert)))$;\\
						\For{$j=1,...,l$}{
							$\mathbf{Q}_{\text{dir}} \gets {\arg\!\min}_{\mathbf{b}_k \in \mathcal{B}_{\rm dir}^{(j)}}  ~d(\tilde{\mathbf{z}}_i,\mathbf{b}_k)$;\\
							$\mathbf{z}^{(j)}_{q,i} \gets Q_{\text{mag}} \cdot \mathbf{Q}_{\text{dir}}$;\\
						}
					}
					$\mathbf{\hat H}^{(j)} \gets f_{\text{dec}}(\mathbf{z}_q^{(j)})$, {$\forall j\in\{1,...,l\}$};\\
					
					Calculate $\mathcal{L}^{(l)}_{\text{nvq}}$ in \eqref{eq: nested quantization};\\
					Update all parameters with the gradient descent;
				}
				$\mathcal{B}_{\text{dir}}^{(l+1)} \gets$ $2^{B_{\text{dir}}^{(l+1)}}$ most quantized vectors in $\mathcal{B}_{\text{dir}}^{(l)}$;
			}
			
	\end{algorithm}}

	
	
	
	\section{Simulation Results}\label{Sec:Simul}
	In this section, we evaluate the performance of the proposed finite-rate CSI feedback method. 
	In our proposed model, network structure in \cite{VQ-VAE} is considered.
	The channel dataset is generated using the COST2100 channel model for the indoor picocellular scenario at 5.3 GHz and outdoor rural scenario at the 300 MHz \cite{COST2100}.
	The BS is equipped with a uniform linear array (ULA) with $N_t = 32$ and $N_c=1024$ in spatial-frequency domain. After transformation to angular-delay domain, we truncate the CSI image with $\tilde{N}_c=32$. The other parameters are described as the default setting in \cite{COST2100}. The dataset contains the training, validation, and testing sets with the size of $100000$, $30000$, and $20000$ respectively. Also, the batch size is set to 200. The Adam optimizer with 0.001 learning rate, and 1000 training epochs are used. 
	For performance comparison, we consider (i) the original VQ-VAE method in \cite{VQ-VAE}, (ii) the DL-based CSI feedback method in \cite{CsiNet+}, (iii) the DL-based CSI feedback method in \cite{test_channel}, and (iv) the DL-based CSI feedback method in \cite{quantization adaptor}.
	The parameters related to the quantization process are set as $A=0.6$, $B_{\text{mag}}=4$, $D=16$, $\mu=255$, $\beta=0.25$, and $\tau=8$.
	A performance measure considered in our simulations is normalized mean squared error (NMSE) defined as $\text{NMSE}=\mathbb{E}\big\{{\big\Vert\tilde{\mathbf H}_{\text{ad}}-\hat{\mathbf H}\big\Vert^2_{\text F}}/{\big\Vert\tilde{\mathbf H}_{\text{ad}}\big\Vert^2_{\text F}}\big\}$.

\begin{figure}[t]
			\centering
			{\epsfig{file=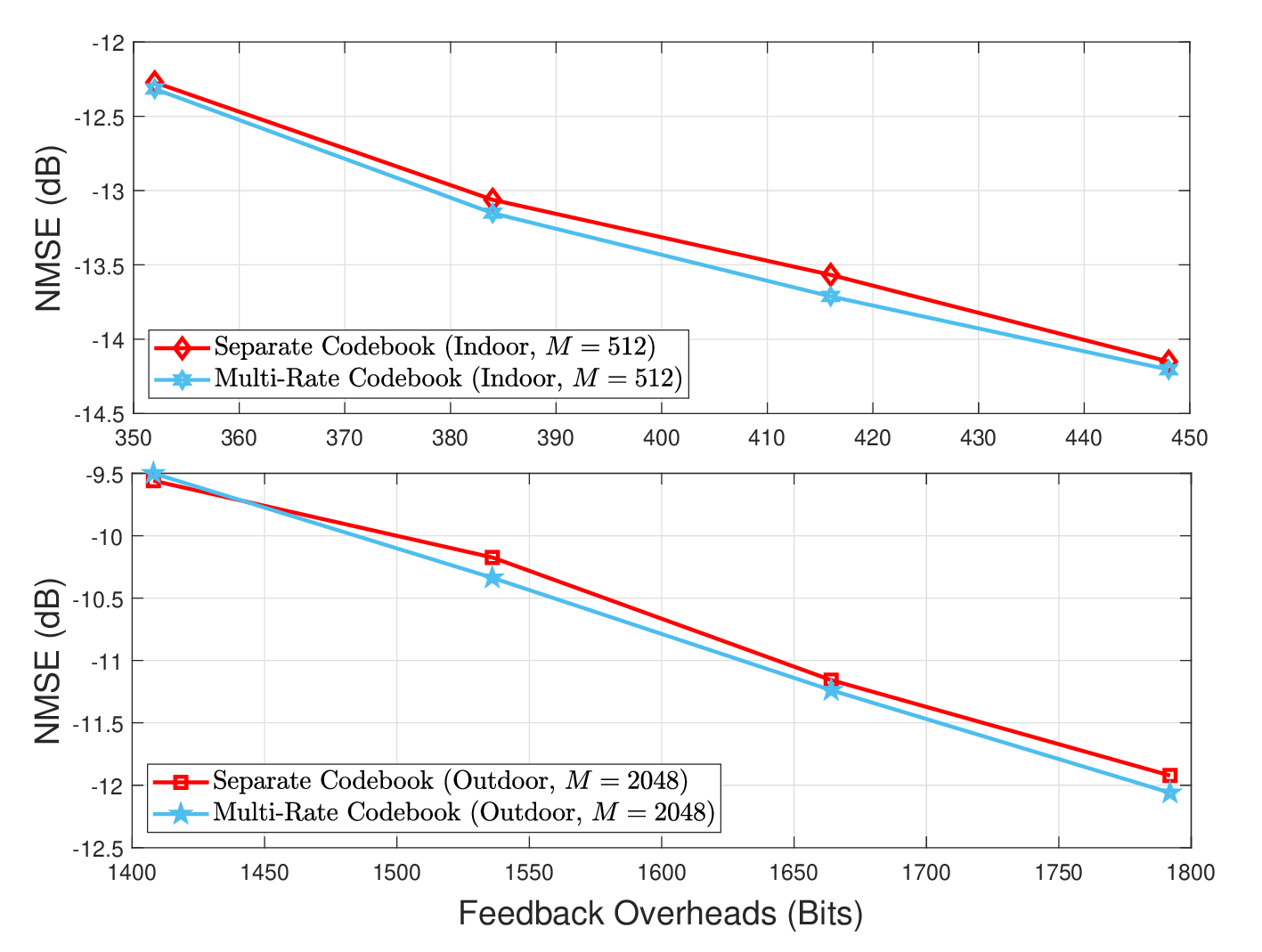, width=10cm}}\vspace{-3mm} 
			\caption{Comparison of the NMSE performance of the proposed method with and without the multi-rate codebook design when $L=4$.} 
			\label{fig:nested_codebook}\vspace{-3mm}
		\end{figure}

		Table I compares the NMSE performance of various CSI feedback methods across different feedback overheads. For the proposed method with multi-rate codebook design, we set $L=2$ and $\gamma=0.8$, indicating that a single nested codebook covers two different rates. Table I demonstrates that, for the same feedback overhead, the proposed method outperforms other CSI feedback methods. This result indicates that the proposed method effectively enhances the performance of the original VQ-VAE approach through a judicious design of the vector codebook. Furthermore, it is shown that the proposed method with multi-rate codebook design achieves better performance than the proposed method with single-rate codebook design. This result demonstrates that our multi-rate codebook design strategy not only reduces the number of required codebooks but also facilitates the effective design of the codebook. A similar result is also depicted in Fig.~\ref{fig:nested_codebook}, which compares the NMSE performance of the proposed method with and without multi-rate codebook design when $L=4$. Fig.~\ref{fig:nested_codebook} shows that the proposed method with multi-rate codebook outperforms the proposed method with the separate design of the codebooks.
		
		\begin{table}[t]
			\label{table: NMSE performance}
			\caption{Comparison of the NMSE ({\em dB}) performance of various CSI feedback methods across different feedback overheads.}
			\centering
			\begin{tabular}{|ccccc|}
				\hline
				\multicolumn{5}{|c|}{Indoor} \\ \hline
				\multicolumn{1}{|c|}{Feedback Overheads (Bits)} & 384 & 512 & 640 & 768 \\ \hline
				\multicolumn{1}{|c|}{Model in {[}3{]} ($M=128$)} & -10.24 & -12.58 & -13.64 & -14.02 \\ \hline
				\multicolumn{1}{|c|}{\begin{tabular}[c]{@{}c@{}}Model in {[}16{]}\\ ($M=128,256,-,256$)\end{tabular}} & -10.72 & -11.71 & - & -13.14 \\ \hline
				\multicolumn{1}{|c|}{VQVAE ($M=1024$)} & -12.31 & -14.17 & -14.63 & -15.21 \\ \hline
				\multicolumn{1}{|c|}{\begin{tabular}[c]{@{}c@{}}Proposed (Single-rate, $L=1$) \\ ($M=512, 512, 1024, 1024$)\end{tabular}} & \textbf{-13.06} & \textbf{-14.79} & \textbf{-15.33} & \textbf{-16.33} \\ \hline
				\multicolumn{1}{|c|}{\begin{tabular}[c]{@{}c@{}} Proposed (Multi-rate, $L=2$) \\ ($M=512\ |\ M=1024$)\end{tabular}} & \textbf{-13.15} & \multicolumn{1}{c|}{\textbf{-14.93}} & \textbf{-15.5} & \textbf{-16.41} \\ \hline
				\multicolumn{5}{|c|}{Outdoor} \\ \hline
				\multicolumn{1}{|c|}{Feedback Overheads (Bits)} & 768 & 1024 & 1536 & 2048 \\ \hline
				\multicolumn{1}{|c|}{\begin{tabular}[c]{@{}c@{}}Model in {[}3{]}\\ ($M=256,256,512,512$)\end{tabular}} & -6.67 & -7.94 & -9.96 & -11.54 \\ \hline
				\multicolumn{1}{|c|}{\begin{tabular}[c]{@{}c@{}}Model in {[}17{]}\\ ($M=128,256,256,512$)\end{tabular}} & -5.51 & -8.10 & -8.35 & -12.13 \\ \hline
				\multicolumn{1}{|c|}{\begin{tabular}[c]{@{}c@{}}VQVAE\\ ($M=2048,2048,3072, 4096$)\end{tabular}} & -6.92 & -7.32 & -9.56 & -10.98 \\ \hline
				\multicolumn{1}{|c|}{\begin{tabular}[c]{@{}c@{}}Proposed (Single-rate, $L=1$) \\ ($M=1024,1024,2048, 2048$)\end{tabular}} & \textbf{-7.11} & \textbf{-9.06} & \textbf{-10.17} & \textbf{-12.7} \\ \hline
				\multicolumn{1}{|c|}{\begin{tabular}[c]{@{}c@{}} Proposed (Multi-rate, $L=2$) \\ ($M=1024\ |\ M=2048$)\end{tabular}} & \textbf{-7.34} & \multicolumn{1}{c|}{\textbf{-9.15}} & \textbf{-10.32} & \textbf{-13.13} \\ \hline
			\end{tabular}
			\vspace{-3mm}
		\end{table}

		
\begin{figure}[t]
			\centering
			{\epsfig{file=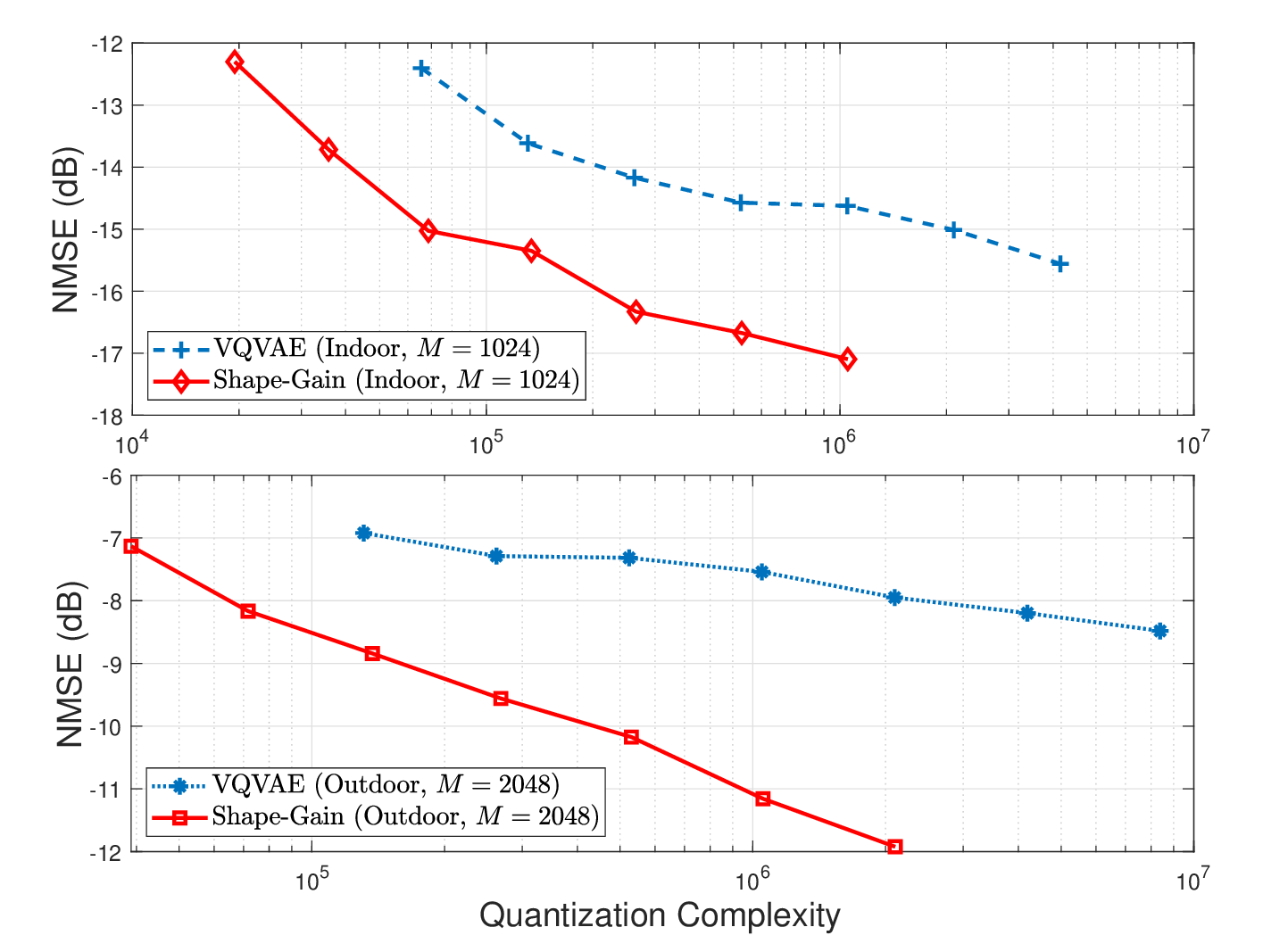, width=10cm}}\vspace{-3mm}
			
			\caption{Comparison of the performance-complexity trade-off achieved by the proposed and original VQ-VAE methods.
			} \vspace{-3mm}
			\label{fig:performance_complexity}
		\end{figure}

		Fig.~\ref{fig:performance_complexity} compares  the trade-off between performance and complexity achieved by the proposed and original VQ-VAE methods.  In this simulation, the computational complexity is measured by the number of multiplications in the quantization process. Fig.~\ref{fig:performance_complexity} demonstrates that, for the same complexity, the proposed  method significantly outperforms the  VQ-VAE method. This result verifies that the proposed method is a powerful solution to improve the performance-complexity trade-off in finite-rate CSI feedback. 
		

		\section{Conclusion}
		
		In this paper, we have presented a finite-rate DL-based CSI feedback method for massive MIMO systems, utilizing the VQ-VAE framework. By leveraging shape-gain vector quantization, we have successfully alleviated the computational complexity associated with VQ-VAE, while improving its reconstruction performance under a given feedback overhead. We have also demonstrated that the presented method can support multi-rate vector quantization by harnessing the principle of the nested quantization.


	\end{document}